\documentclass[twocolumn,aps, prl,showpacs]{revtex4}

\usepackage{graphics}
\usepackage{graphicx}
\usepackage{dcolumn}
\usepackage{bm}
\usepackage{amsmath,amssymb}
\usepackage{color}
\usepackage{soul,xcolor}
\setstcolor{red}

\newcommand{\grad}[1]{\nabla #1}

\begin{document}

\title{Optical analogues of the Newton-Schr\"odinger equation and boson star evolution}
\author{Thomas Roger$^{1}$, Calum Maitland$^{1}$, Kali Wilson$^{1}$, Niclas Westerberg$^{1}$, David Vocke$^{1}$, Ewan M. Wright$^{2,1}$, Daniele Faccio$^{1}$}
\email{d.faccio@hw.ac.uk}
\affiliation{$^{1}$School of Engineering and Physical Sciences, Heriot-Watt University, EH14 4AS Edinburgh, UK\\
$^{2}$College of Optical Sciences, University of Arizona, Tucson, USA}

\date{\today}
\begin{abstract}
{
Many gravitational phenomena that lie at the core of our understanding of the Universe have not yet been directly observed. An example in this sense is the boson star that has been proposed as an alternative to some compact objects currently interpreted as being black holes. In the weak field limit, these stars are governed by the Newton-Schrodinger equation (NSE). Here we present an optical system that, under appropriate conditions, identically reproduces the NSE in two dimensions. A rotating boson star is experimentally and numerically modelled by an optical beam propagating through a medium with a positive thermal nonlinearity and is shown to oscillate in time while also stable up to relatively high densities. For higher densities, instabilities lead to an apparent breakup of the star, yet coherence across the whole structure is maintained. These results show that optical analogues can be used to shed new light on inaccessible gravitational objects.
}
\end{abstract}
\pacs{42.65.Tg, 42.65.Jx, 42.65.-k, 42.65.Sf}
\maketitle
\noindent {\bf Introduction}\\
Analogue gravity is the study of gravitational effects using artificial systems that recreate some specific aspects of the full gravitational system using laboratory-based experiments \cite{review}. Recent experimental studies have focused attention on analogues for black holes and the search for Hawking radiation in transonic fluid flows \cite{unruh} that may be obtained in a variety of systems such as flowing water \cite{Weinfurtner}, Bose-Einstein condensates \cite{Lahav,Steinhauer} and nonlinear optics \cite{Leonhardt,Faccio,Faccio2}.  To date, attempts to experimentally study gravitational systems using analogues have tended to focus on the local and linear dynamics, i.e. on the behaviour of weak perturbations or waves in background curved spacetime geometry, determined solely by the local flow speed of the supporting fluid or medium. However, gravity is inherently a nonlinear and nonlocal, i.e. long-range, interaction. This has been used to draw an analogy between gravitational attraction and light-trapping in the wake of optical solitons \cite{Gorbach}. Another notable example of the connection between gravity and optics was recently demonstrated using an optical system based on a thermally excited medium, which allows one to reproduce the physics of the Newton-Schr\"odinger equation (NSE) \cite{Bekenstein2}. The NSE can be written as 
\begin{equation}
i\hbar\psi_t + \frac{\hbar^2}{2m}\grad^2\psi+m\phi\psi = 0\text{,}
\label{eqn:NSE}
\end{equation}
which describes the evolution of a particle with mass $m$ defined by a wave function $\psi$. $\grad^2$ is the three-dimensional Laplacian and the subscript $t$ denotes a derivative with respect to time. The gravitational potential $\phi$ is determined by the Poisson equation
\begin{equation}
\grad^2\phi = -4\pi G m|\psi|^2\text{,}
\label{eqn:curvG}
\end{equation}
where $G$ is the gravitational constant. This equation was proposed by Di\`osi \cite{diosi} and Penrose \cite{penrose} as an attempt to investigate quantum wave function collapse in the presence of a Newtonian gravitational potential. More recently, the same equation has been used, as in its original proposal \cite{ruffini} to describe the evolution of boson or Bose-Einstein condensate stars. Indeed, the NSE may be obtained as the non-relativistic limit of the Klein-Gordon equation \cite{giulini}, and describes the coupling of  classical gravitational fields to quantum matter states. Importantly, and in departure from most previous ``linear'' analogue gravity studies, here the gravitational potential includes the quantum matter itself as a source. The remarkable feature that allows one to build laboratory experiments is that the evolution of the amplitude $\cal E$ of an optical beam in a thermally focusing medium is described by  
\begin{equation}
i{\partial{\cal E}\over\partial z} + {1\over 2k} \nabla_\perp^2{\cal E}+k_0\Delta n {\cal E}=0 ,
\label{eqn:NLSE}
\end{equation}
where $ \nabla_\perp^2$ is the transverse, two-dimensional Laplacian, $k={n_b\omega/ c}=n_bk_0$, $n_b$ being the background refractive index and the nonlocal change in refractive-index, $\Delta n$ is induced by heating of the medium by the beam itself
\begin{equation}\label{Tss}
\nabla_\perp^2 (\Delta n) = -{\alpha\beta\over\kappa} |{\cal E}|{^2}  , 
\end{equation} 
where $\beta=\partial n/\partial T$ is the medium thermo-optic coefficient,  $\kappa$ is the thermal conductivity and $\alpha$ the absorption coefficient. \\
The similarity between the gravitational equations \eqref{eqn:NSE} and \eqref{eqn:curvG} and the nonlinear optical beam propagation equations \eqref{eqn:NLSE} and \eqref{Tss} indicates the possibility to simulate a 2D slice of the full 3D gravitational system. We note that in the optical model, the propagation of the optical beam through space ($z$) maps onto the time coordinate of the NSE. 

\begin{figure}[t!]
\begin{centering}
\includegraphics[width = 0.5\textwidth]{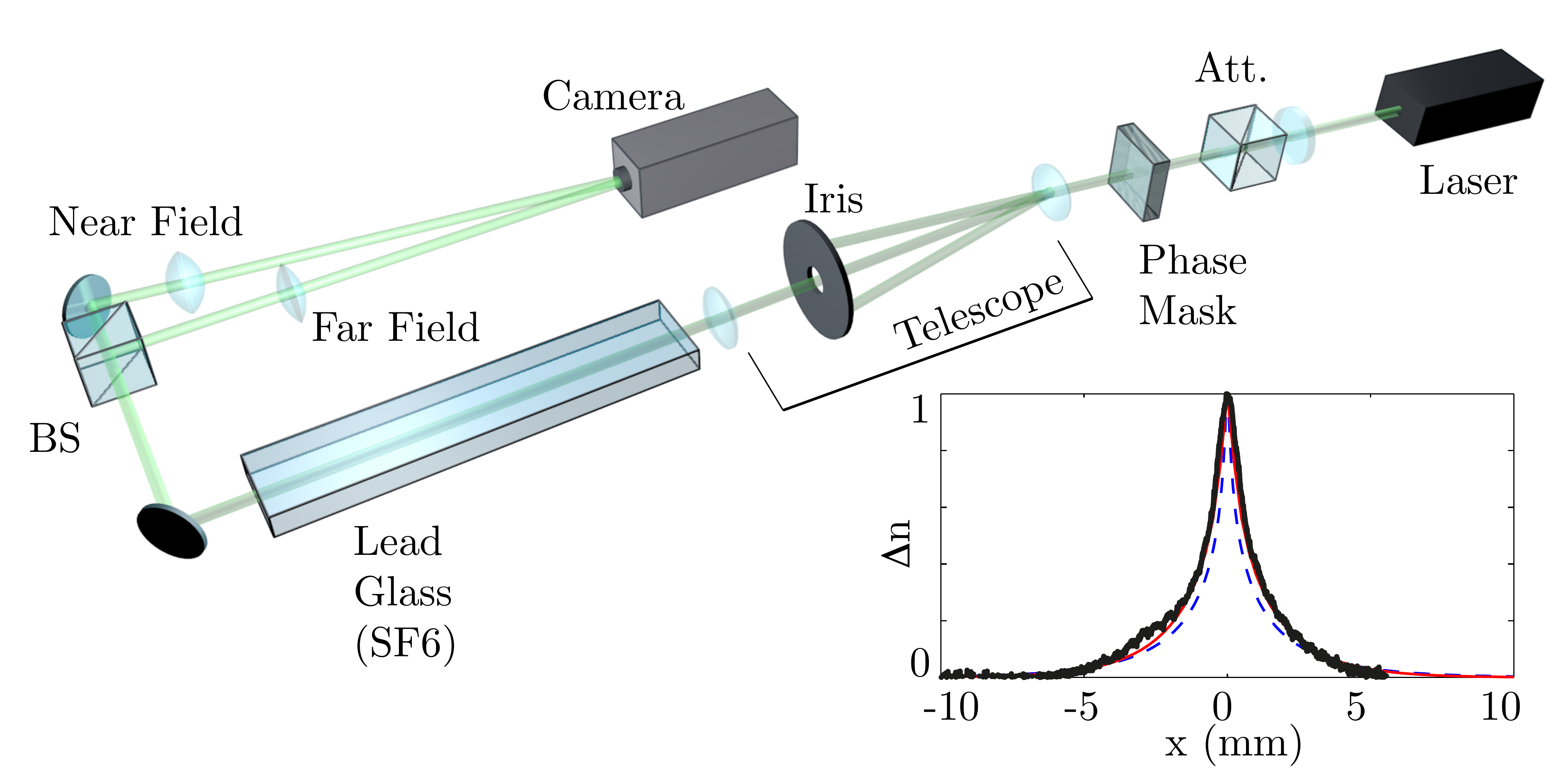}
\caption{\textbf{Experimental layout.} The Newton-Shcr{\"{o}}dinger equation is simulated using a continuous-wave laser centred at 532 nm, which is passed through an attenuator (Att., $\lambda$/2-waveplate + polariser) then through an azimuthal phase mask. The phase mask imprints a vortex phase structure $e^{i\ell\phi}$ creating a series of diffracted orders. The first-order (with $\ell = 1$) is isolated with an iris in the far field of an $f = 1$ m lens and collected with an $f = 250$ mm lens forming a telescope with a $4:1$ de-magnification factor. The vortex beam is centred onto the input facet of a lead-doped glass slab (height D=5 mm, length 400 mm, width 40 mm). The near and far fields of the output facet of the glass slab are imaged separately with two lenses onto a camera so as to monitor the real-space intensity distribution and the spatial-frequency spectrum of the simulated boson star. The inset graph shows the heat-induced refractive index change ($\Delta n$, normalised to one) in the glass sample as predicted by the full numerical solution to Eq.~\eqref{Tss} (blue dashed line), by the distributed loss model (red line, $\sigma=D/2$), and the experimentally measured $\Delta n$ (thick black line).}
\label{fig:expt}
\end{centering}
\end{figure}

\begin{figure}[t]
\begin{centering}
\includegraphics[width = 0.5\textwidth]{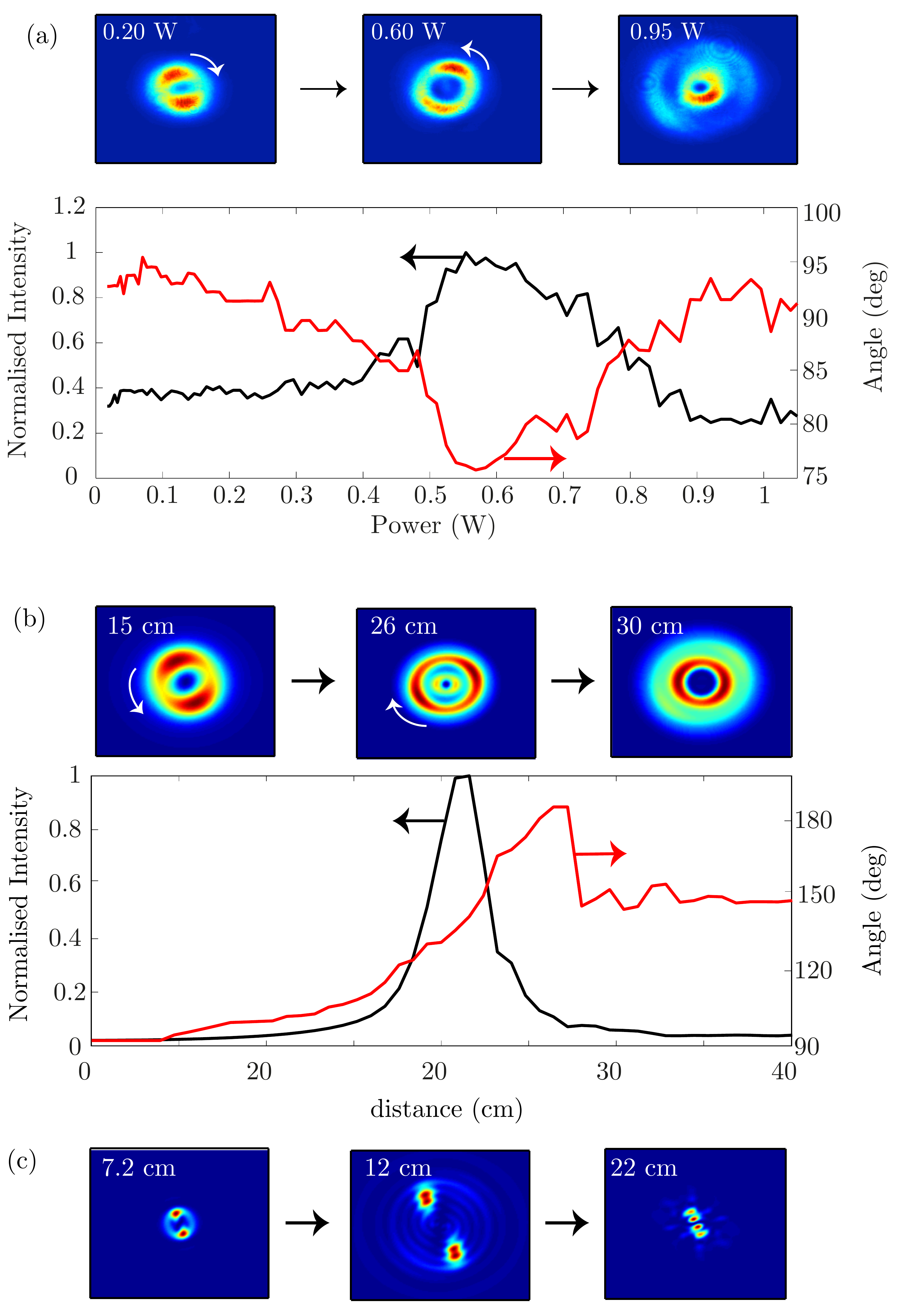}
\caption{\textbf{Transverse plane images of an analogue rotating boson star.} (a) Experiments: the top three panels show examples of the transverse profile of the analogue boson star at different input powers. The graph shows the measured evolution of the peak intensity (left $y$-axis, black trace) versus input power for the experiments shown in Fig. \ref{fig:expt}.  The rotation angle of the azimuthal features is shown for increasing input power on the right $y$-axis (red trace). (b) Numerical simulation  of a vortex beam with a similar spatial profile to those seen in the experiments. The rotation angle of the azimuthal features is shown  for a fixed input power of 1.6 W for increasing propagation lengths. A similar non-monotonic rotation speed of the peaks is observed as in the experiments. (c) Same as in (b) but for a higher input power of 10 W. The transverse profiles show a more complicated evolution, with a break up into multiple intensity peaks, while the overall coherence is maintained, allowing the contraction to be counteracted by the central phase singularity.}
\label{fig:rot}
\end{centering}
\end{figure}

In this article we show that if care is taken in choosing the appropriate operating conditions, the medium response in the transverse plane of the optical beam can be made to mimic exactly the gravitational scenario. More precisely, we find that the exact correspondence is obtained when the optical beam carries no energy on-axis in the Fourier domain, i.e. when the pump is ring-shaped in $K$-space. We then use this correspondence to experimentally study the evolution of a rotating boson star that has quantised angular momentum \cite{Yoshida,Liebling} using an intense light beam propagating in a lead-glass slab. For low star densities (beam intensities) the evolution is characterised by self-focusing contraction cycles. The experimental results are well reproduced by numerical simulations, which are then extended to higher densities than those achievable in the experiments. At these high densities, the star becomes unstable yet complete collapse is prevented by the phase singularity (related to the quantised angular momentum of the entire structure) at the star centre. \\ \newline
{\bf Results}\\
\noindent {\bf Theoretical Model.}
The basic model for an optical beam propagating in a thermo-optic medium is derived within the paraxial approximation and incorporates the thermally-induced change in refractive index $\Delta n({\bf r}_\perp,z)$, ${\bf r}_\perp=(x,y)$ being the transverse position vector.  The resulting paraxial wave equation for the slowly-varying complex electric field envelope propagating along the z-axis is
\begin{equation}
{\partial{\cal E}\over\partial z} = {i\over 2k} \nabla_\perp^2{\cal E}+ik_0\Delta n {\cal E} - {\alpha\over 2}{\cal E} ,
\end{equation}
where $\nabla_\perp^2$ is the transverse Laplacian, $\alpha$ is the medium absorption coefficient, and the nonlocal change in refractive index is expressed by the integral \cite{Minovich,Vocke}
\begin{equation}\label{deln}
\Delta n({\bf r}_\perp,z) = \gamma \int_{ }^{ } d^2{\bf r}_\perp'~R({\bf r}_\perp;{\bf r}_\perp') I({\bf r}_\perp',z), 
\end{equation} 
where $I({\bf r}_\perp,z)=|{\cal E}({\bf r}_\perp,z)|^2$ describes the beam intensity profile, and $\int_{ }^{ } d^2{\bf r}_\perp~R({\bf r}_\perp;{\bf r}_\perp')  = 1$.
Thus in the case of a local nonlinear optical response, $R({\bf r}_\perp; {\bf r}_\perp')=\delta({\bf r}_\perp -{\bf r}_\perp')$, the nonlinear coefficient is given by $\gamma$. In the above equations, $R({\bf r}_\perp;{\bf r}_\perp')$ is the response function for the medium, which incorporates the transverse boundary conditions. We assume the transverse boundary does not vary with $z$, and that the medium is sufficiently long, and optical absorption low enough that the thermal diffusion is dominated by transverse diffusion and longitudinal effects are not significant. \\
\newline
\noindent {\bf Response function.} It is by tailoring the medium response function that one may ultimately reproduce a precise analogue of the NSE in the laboratory. Physically the change in refractive index in the medium arises from the temperature increase due to laser absorption $\Delta n=\beta\Delta T$, with $\beta={\partial n\over\partial T}$.  Taking into account appropriate boundary conditions, the spatial profile of the temperature change obeys the heat equation 
\begin{equation}
(\rho_0 C){\partial\Delta T\over \partial t} = \kappa\nabla_\perp^2 (\Delta T) + \alpha I,
\end{equation}
with $\kappa$ the thermal conductivity, and $\rho_0 C$  the heat capacity per unit volume.  Finally in steady state, found by setting the time derivative to zero, we obtain Eq.~\eqref{Tss} 
which should be solved subject to the appropriate boundary conditions due to heat loss. Equation (\ref{deln}) shows that the response function is, to within a constant, the Green's function for heat diffusion in the medium, and obeys
\begin{equation}\label{Req1}
\nabla_\perp^2 R({\bf r}_\perp; {\bf r}_\perp') =-\left ({\alpha\beta\over\kappa\gamma}\right ) \delta({\bf r}_\perp - {\bf r}_\perp').
\end{equation}
\newline
\noindent {\bf Infinite space model.}
In this case the boundaries are moved off to infinity in the two transverse dimensions, and we might hope to have a realization of the Schr\"odinger-Newton equation albeit in 2D.   Fourier transforming Eq. (\ref{Req1}) yields 
\begin{equation}
\tilde R({\bf K}_\perp) = \left ({\alpha\beta\over\kappa\gamma}\right ){1\over {\bf K}_\perp^2}  ,
\end{equation}
where the tilde signifies a transformed variable, and ${\bf K}_\perp=\sqrt{K_x^2+K_y^2}$. The same functional form for $\tilde R$ is of course also expected for the gravitational case. However, in the two dimensions used in our experiments (the transverse plane of our optical beam), the resulting response function in the real-space domain is
\begin{equation}
R({\bf r}_\perp - {\bf r}_\perp') = -\left ({\alpha\beta\over\kappa\gamma}\right ){1\over 2\pi}\ln\left (|{\bf r}_\perp - {\bf r}_\perp'|\right ) ,
\end{equation}
which is very different from the Coulomb-like (decaying) interactions that arises in three dimensions and in the gravitational case.  \\ \newline
\noindent {\bf Distributed loss model.} The lesson from the infinite space model is that thermal losses due to the presence of the boundaries will always have to be incorporated at some level.   We consider the case for which the boundaries are present but well removed by a characteristic distance $D/2$ from the transverse position at which the laser beam of width $W\ll D/2$ is centered.  For a cylindrical medium, $D$ will be the medium diameter, whereas for a medium of rectangular cross section, $D$ is chosen as the smaller dimension.  Then to a reasonable approximation the medium will be shift-invariant for displacements in the vicinity of the beam center, i.e. the boundaries should not significantly impact upon the symmetry of the laser beam as it propagates.  In this regime we may include the effect of the distant boundaries into a distributed loss term in the starting temperature equation
\begin{equation}
(\rho_0 C){\partial\Delta T\over \partial t} = \kappa\nabla_\perp^2 (\Delta T) + \alpha I - {\kappa\over\sigma^2} \Delta T .
\end{equation} 
Here $\sigma$ is a nonlocal interaction length scale set by thermal diffusion and can be taken as $\sigma \sim D/2$. 
Indeed, we measured the nonlocal length using the method put forward by Minovich \emph{et al.} \cite{Minovich}, providing evidence that to a good approximation $\sigma=D/2$ in our system, as shown in Fig.~\ref{fig:expt}. Numerical solutions to the thermal diffusion equation, indicate that this is a generic feature and not a peculiarity of our system. This type of thermal nonlocality was also previously explored in the context of vortex solitons in Ref.~\cite{YakZalKiv05}.  
With the addition of the distributed loss term, the response function in Eq.~(\ref{Req1}) is described by
\begin{equation}\label{Req2}
\left [ \nabla_\perp^2-{1\over\sigma^2}\right ] R({\bf r}_\perp-{\bf r}_\perp')=-\left ({\alpha\beta\over\kappa\gamma}\right ) \delta({\bf r}_\perp - {\bf r}_\perp'),
\end{equation}
and by Fourier transforming we obtain
\begin{equation}\label{RK1}
\tilde R({\bf K}_\perp) = \left ({\alpha\beta\sigma^2\over\kappa\gamma}\right ){1\over (1+\sigma^2 {\bf K}_\perp^2) }  .
\end{equation}
The Fourier transform of Eq.~(\ref{deln}) yields $\Delta\tilde n({\bf K}_\perp) = \gamma \tilde R({\bf K}_\perp) \tilde I({\bf K}_\perp)$ and the effective nonlinear coefficient is given by \cite{Boyd}
\begin{equation}\label{gamma}
\gamma = \left ({\alpha\beta\sigma^2\over\kappa}\right ) .
\end{equation}
We finally obtain a  relation for the $K$-space response function
\begin{equation}\label{RK}
\tilde R({\bf K}_\perp) = {1\over 1+(\sigma {\bf K}_\perp)^2 },
\end{equation}
and the corresponding response function in real space, 
\begin{equation}
R({\bf r}_\perp-{\bf r}_\perp')={1\over 2\pi\sigma^2} K_0\left ({|{\bf r}_\perp-{\bf r}_\perp'|\over\sigma} \right ),
\end{equation}
where $K_0$ is the zeroth-order modified Bessel function of the second kind. 
 In our experiments we use a similar lead-doped glass to that used in Ref.~\cite{RotCohMan05} with parameters $\kappa=0.7$ Wm$^{-1}$K$^{-1}$, $n_b=1.8$, $\beta=14\times 10^{-6}$ K$^{-1}$, and $\alpha = 0.01$ cm$^{-1}$.  In our experiments described below $D=5$ mm, giving $\sigma = 2.5$ mm and $\gamma=1.25\times 10^{-6}$ cm$^2$/W.\\ \newline
\noindent {\bf Comparison of models and reduction to the NSE.}  We start by noting that when the longitudinal wavevector component $K_z\ll {\bf K}_\perp$, the Coulomb response function in the NSE in K-space reduces to $\tilde R({\bf K}_\perp) =  {1/{\bf K}_\perp^2}$. 
We then recall that for the infinite space model, and using Eq.~(\ref{gamma}), the Fourier transform of the response function may be written as $\tilde R({\bf K}_\perp) =  {1/(\sigma {\bf K}_\perp)^2}$, which indeed reproduces the desired Coulomb interaction however, at the expense of a non-physical logarithmic behaviour in real space. The latter is due to the lack of boundaries whose effect is correctly modelled in the `distributed loss model' approach, which provides an exponentially decaying behaviour in direct space. In K-space, the `distributed loss model' response function reduces to the Coulomb form in the limit $(\sigma {\bf K}_\perp)^2\gg1$. 
Therefore, for a general input beam (typical laser beams have a Gaussian profile) whose spatial-frequency spectrum includes the region around the origin ${\bf K}_\perp=0$, the  thermal and the NSE models will be significantly different.  However, if we use beams comprising wavevectors only of $|{\bf K}_\perp|\gg1/\sigma$, then we may realize the dynamics of the infinite-space system, and therefore also of the gravitational system, while still accounting for the boundaries as above. With this caveat, our 2D thermo-optic model has the same response function as the gravitational Poisson equation used in the NSE, thus providing the basis for the correspondence between our nonlocal optical and the Newtonian gravitational system. \\
Finally, we point out that the propagation along the $z$-axis of our optical beam is analogous to propagation in time of a particle in the NSE framework. Furthermore,  a gravitational-like potential requires that the sign of $\Delta n$ is positive. \\
\newline
\noindent {\bf Boson stars.}  We may now use this model to perform experiments that reproduce the dynamics of the NSE. In particular, we chose to study a specific astrophysical object that is described by the NSE, namely a boson star. Originally envisioned by John Wheeler in the 1950s as localised bundles of electromagnetic energy named \emph{geons} \cite{Power}, boson stars are the stable particle-like solution to the Klein-Gordon equation \cite{Kaup, Schunck}. Although boson stars may or may not exist in nature they provide a useful testbed for the study of compact objects that can be described by a single wave function. Indeed, boson stars are assumed to be described by a wave function, which is governed by the Einstein-Klein-Gordon (EKG) equations \cite{Liebling}. The star then arises as a result of a balance between the gravitational field and the dispersion of the wave function. This dispersion essentially originates from the Heisenberg uncertainty principle, i.e. the star counteracts gravity due to the impossibility to arbitrarily localise the star's wave function in both position and momentum.  In this work we consider a specific type of rotating boson star which is taken in the weak field (Newtonian) limit where the EKG equations reduce to the NSE \cite{Guzman}. The fact that a boson  star is described by a single-valued wave function implies that  the angular momentum must be quantised. More explicitly, the wave-function is of the form $\psi(r,t,\theta)=|\psi(r,\theta)|e^{i(\omega t + m\theta)}$, where $\psi(r,t,\theta)=\psi(r,t,\theta+2\pi)$ implies that the angular momentum number $m$ is an integer  \cite{Liebling,rot1,rot2, rot2D,rot3,rot4,rot5}. This in turn leads to a toroidal mass distribution in the star, which can be seen as a result of the fact that the phase is undefined at the star centre \cite{rot1,rot2}. These features hold in both 3D and 2D \cite{rot2D} and emerge naturally in optical experiments using beams that carry orbital angular momentum. \\ \newline
\noindent {\bf Experiments.} A continuous-wave laser with central wavelength 532 nm is used to pump a slab of lead-doped glass with positive thermo-optic nonlinearity (see Methods). We imprint an azimuthal phase onto the Gaussian output beam of the laser via a transmissive phase mask in order to provide the large wavevectors required to operate in the region $(\sigma {\bf K}_\perp)^2 \gg 1$. The beam, now carrying orbital angular momentum (OAM), is imaged after passing through the nonlinear medium onto a CCD camera (see Fig. \ref{fig:expt}). We monitor both the near and far-field of the output facet of the medium while changing the input power of the beam, $P$. The near field provides the real space dynamics $I({\bf r}_\perp, P)$ of the boson star while the far-field reveals the spatial-frequency spectrum, $I({\bf K}_\perp, P)$. We change the power in the absence of being able to track the beam profile as a function of the propagation distance $z$. 
\begin{figure}[h!]
\begin{centering}
\includegraphics[width = 0.5\textwidth]{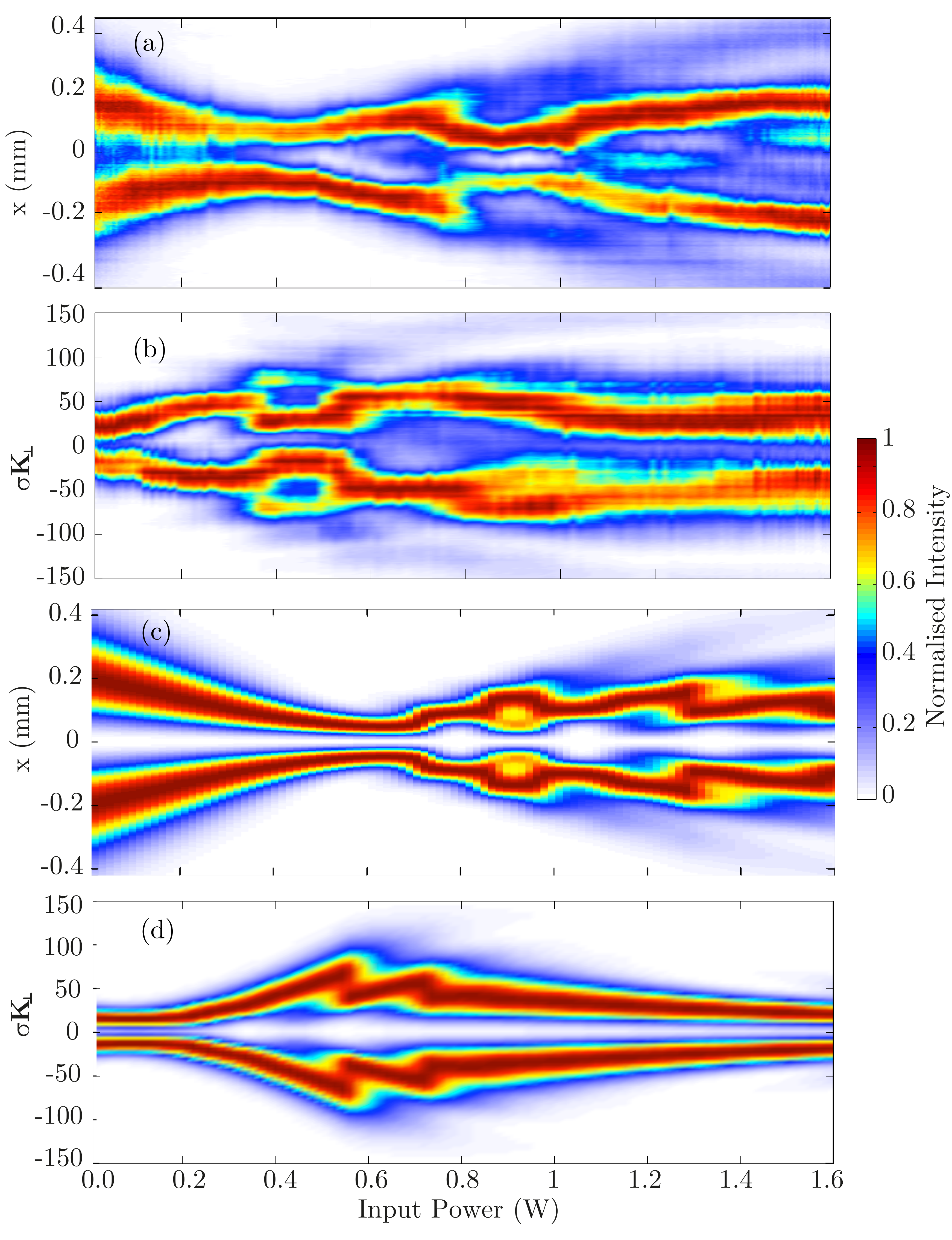}
\caption{\textbf{Experiments and numerics studying the dynamics of analogue rotating boson stars.} (a) The real-space intensity distribution $I(x, 0, P)$ of a vortex beam with $\ell = 1$ over a propagation distance of 400 mm maps the stable time evolution of a rotating boson star. (b) The spatial-frequency spectrum is centred around a value $\sigma {\bf K}_\perp \sim20$, and importantly does not contract to a value smaller than this at any point during the evolution. (c, d) show the numerical simulation of boson star evolution, where (c) is the real-space intensity distribution and (d) the product of the nonlocal length and transverse ${\bf K}_\perp$-vector, $\sigma {\bf K}_\perp$. While (c) confirms that varying the power reasonably reproduces the evolution as a function of distance, (d) shows that throughout the entire evolution of the vortex beam, $\sigma {\bf K}_\perp$ remains sufficiently large to reproduce the NSE and therefore simulate a rotating boson star. }
\label{fig:res_expt}
\end{centering}
\end{figure}
\indent At very low beam intensities we see a ring shaped distribution in both the near and far-fields. Figure \ref{fig:res_expt}(a) shows a 2D slice in the $x$-plane of the near-field evolution $I(x, 0, P)$ of the vortex beam as the pump power is increased from zero  to a maximum power at the sample input facet of $\sim$1.6 W. The beam starts to self-focus as the power is initially increased, reaching a minimum ring diameter at $\sim0.4$ W. The ring then expands before collapsing again at $\sim$0.8 W. It can be seen that there is minimal transfer of energy to higher order modes suggesting that the energy is predominantly localised within the ring. The far-field evolution is shown in Fig. \ref{fig:res_expt}(b), which at low intensities has a spectrum centred on a ring with $\sigma {\bf K}_\perp \approx 20$. As the far-field spectrum and near-field intensity distribution are inversely linked through their Fourier transform we see that the far-field expands at first and contracts after the near-field passes through the first minimum at $\sim 0.4$ W. There is some indication of mode splitting hereafter but the majority of the energy is conserved within the single input mode. We note that throughout the entire propagation the spectrum does not contract to a value smaller than the initial $\sigma {\bf K}_\perp$, thus ensuring that at all times we are  in a regime [$(\sigma {\bf K}_\perp)^2\gg 1$] under which the analogy with the NSE equation is valid. 
The self-contraction cycles in the optical system are due to the interplay between nonlinearity-induced self-focusing and saturation mechanisms that prohibit the beam to continue collapsing indefinitely. Other optical systems are known to exhibit similar dynamics, e.g. laser pulse filamentation where the saturation and counterbalancing effect is provided by plasma defocusing (see e.g. \cite{Couairon}) . Here the contraction can be arrested by two mechanisms: the first is diffraction, analogous to the repulsive pressure term due to the uncertainty principle in the gravitational case. This will occur even in the absence of angular momentum. The second is due to the presence of a phase singularity at the centre of the beam (due to the quantised angular momentum) that enforces zero-light intensity there.  For gravitational Boson stars, numerical studies have shown similar contraction cycles in the absence of angular momentum (\cite{Liebling,Lai}). \\ \newline
\noindent {\bf Numerical simulations.} We have performed simulations based on Eqs. (5) and (6) using the split-step beam propagation method \cite{FMF}.  In this approach propagation over a length $L$ is written as a product of $N_z$ small steps $\Delta z=L/N_Z$.  For large enough number of longitudinal sample points $N_z$  this allows the propagation to be approximated as a concatenation of small propagation steps of length $\Delta z$ involving only the diffraction term on the right-hand-side of Eq. (5),  followed by the same but involving only the nonlinear term.  The small step of diffraction is performed in the transverse spatial frequency domain using fast-Fourier transforms on a transverse grid of $N_T\times N_T$ points.  The nonlinear refractive-index in Eq. (6) is evaluated by converting the convolution to a product in the spatial frequency domain and, using Eq. (15), the resulting small nonlinear step is performed  point-by-point over the transverse plane.  For the simulations presented we have used $N_z=1000$ longitudinal sample points along with $N_T=1024$ transverse grid points, convergence being checked by increasing the number of points. We use the same experimental parameters for the material listed above.
\begin{figure}[t]
\begin{centering}
\includegraphics[width = 0.5\textwidth]{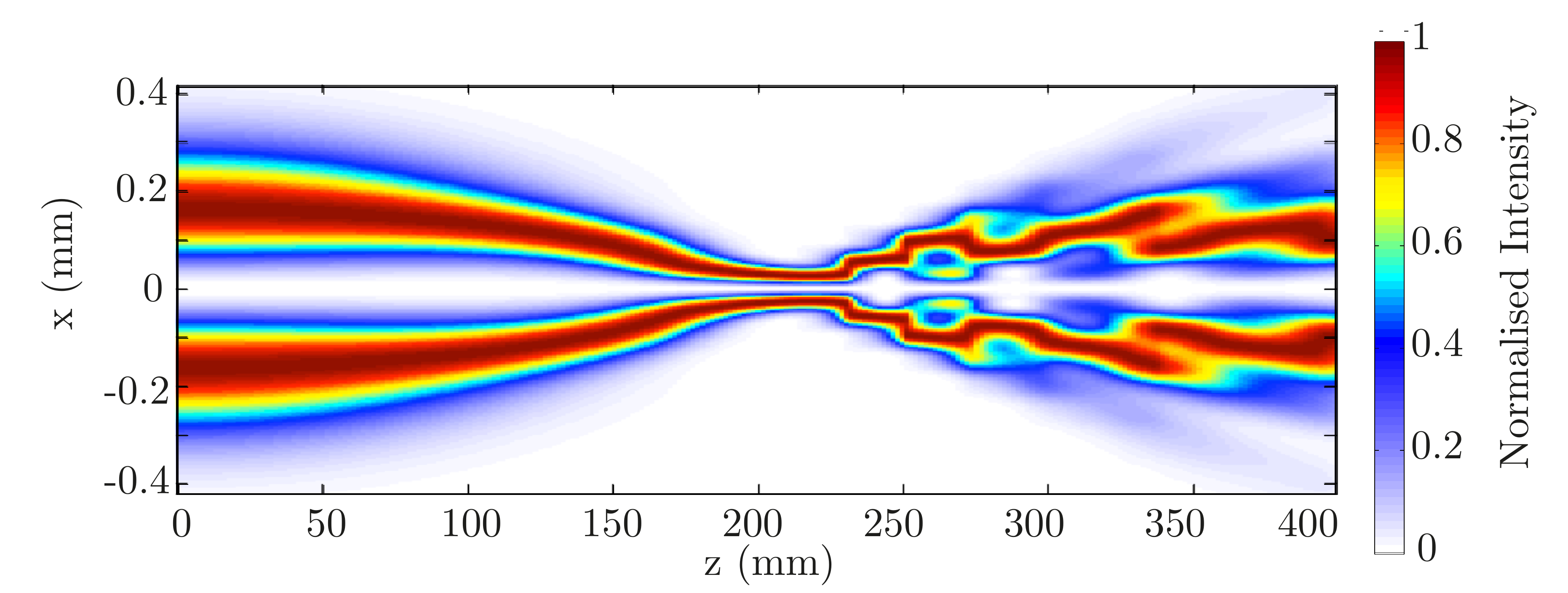}
\caption{\textbf{Numerical simulation of experimental  boson star evolution in time.} We show here the evolution of the optical wave packet according to Eq. (3) In this case the propagation of the optical field in space ($z$) now maps precisely onto the time coordinate of the boson star evolution.} 
\label{fig:res_num}
\end{centering}
\end{figure}
Thus, for the beam we use an input ring diameter $w_0 = 360~\mu$m, a maximum power input of $P = 1.6$ W and 400 mm of propagation distance. Figure~\ref{fig:res_expt} (c-d) shows the near field intensity as a function of the input power. These simulations can then be directly compared to the evolution of optical wavepacket as the power is increased (see Fig.~\ref{fig:expt}(a-b)) with good agreement. \\
In Fig.~\ref{fig:res_num} we show the evolution of the boson star in time (i.e. intensity distribution along the propagation length of the optical beam $I(x,0,z)$). Again, we see an initial contraction of the ring beam followed by a series of oscillations in which the majority of the energy is conserved in the innermost ring, with little lost to higher order modes. \\
\newline
\noindent {\bf Transverse near-field dynamics.} We now consider the transverse profile of the beam. In the experiments we introduce a slight azimuthal anisotropy, which is manifested as two slightly higher intensity peaks which live on symmetrically opposite sides of the ring (see top panels in Fig. \ref{fig:rot}(a)). Figure~\ref{fig:rot}(a) shows the maximum intensity in the beam (left y-axis) and the angle of the peaks as a function of the input power (right y-axis).  The two peaks are seen to slowly rotate around the ring as the beam self-focuses with increasing input power. However, the rotation speed suddenly increases close to the tightest contraction point and then actually inverts direction after this point. 
This behaviour is confirmed by the numerical simulations shown in Fig.~\ref{fig:rot}(b)  that allow us to observe the same rotation dynamics along the actual propagation axis for a fixed input power of 1.6 W. The  non-uniform acceleration of the peaks reveals complicated dynamics for the boson star that has not emerged from previous studies. It is possible to interpret the observed rotation as due to Berry's phase, i.e. a geometric phase acquired by the system as it cycles through a closed loop in parameter space. Indeed, based on an analysis of the corresponding linear case (see Supplementary Discussion) where we predict a similar rotation effect, we note that the rotation of the peaks appears as a consequence of a Berry phase accumulation that occurs through the beam focus \cite{berry1,berry2}. Garrison and Chiao have also shown that geometrical phases will arise in a general class of nonlinear field theories displaying global gauge invariance \cite{berry3}, as in the case investigated here. Thus the peak rotation in our experiments and simulations may be viewed as a nonlinear extension of the linear Berry phase occurring during the self-focusing of the fields  seen in Fig.~\ref{fig:rot}.\\
\indent In Fig.~\ref{fig:rot}(c) we show numerical simulations for a much higher 10 W input power. The boson star is now seen to apparently break up into two well separated peaks as it contracts. However, these peaks still form a single coherent wave function. Indeed, whilst two independent peaks or boson stars will overlap and create interference fringes \cite{Liebling}, here the merging of these two peaks is prohibited by the input phase singularity, which survives the evolution of the star. Another way of saying this is that the phase singularity   acts as a perfectly repulsive point where the momentum of the wave function and thus its kinetic energy diverges to infinity, thus preventing collapse of the star and possibly ultimately limiting the formation of a black hole singularity (see Supplementary Discussion). \\

\noindent {\bf Discussion}\\
\noindent We have shown that, when appropriate attention is paid to boundary conditions, a medium with positive thermal-optical nonlinearity may provide a testbed for simulations of the Newton-Schr\"{o}dinger equation. Using optical vortex beams we are able to map the propagation of the wave function along the optical axis to the time evolution of a rotating boson star. However, a thermally focusing medium is not a ``universal'' analogue of the NSE. The nonlocal nonlinearity needs to be properly tailored, e.g. by tailoring the input beam shape in order to enforce the correspondence. We can therefore simulate only a certain subset of objects in the NSE context and we have identified rotating boson stars as one such example that we believe is also of practical interest given that most stars are indeed expected to be in rotation. Therefore, these experiments provide new possibilities for studies into objects described by the NSE. Future work may concentrate on beams with alternate phase profiles, such as Bessel beams and such experiments and studies in the laboratory-based system can provide new insights and research directions for the astrophysical system. Expanding the analysis to include the next higher order term in the NSE hints that the instabilities observed here will actually be stronger in the full relativistic settings (see Supplementary Discussion). It would therefore be interesting to investigate the full relativistic description of phase singularities in rotating objects and their role in the gravitational collapse of boson-star-like objects. While our optical analogue can reproduce a 2D slice of the full 3D behaviour of the NSE, an interesting direction would be to look at systems capable of simulating the full 3D NSE. We note that the equations that govern rotating 2D optical beams with a phase singularity at the centre are of the same form as those that describe a rotating boson star, whose geometry is a torus with a similar phase singularity at the centre \cite{Liebling}. As our analogue reproduces a slice of this torus we expect that insight gleaned from our experiments should be transferrable to `real world' cosmological objects. \\

\noindent {\bf{Methods.}}\\
\noindent {\bf Experimental details}\\
\noindent Our experiments exploring the dynamics of rotating boson stars use a continuous-wave laser with central wavelength 532 nm to pump a slab of lead-doped glass.  The Gaussian beam is first passed through an optical fork-patterned phase mask which imprints an azimuthal phase $e^{i\ell\phi}$, where  $\ell$ is the diffraction order. In the experiments performed here we use a beam with $\ell = 1$, which generates a ring-shaped beam with $I({\bf K}_\perp=0)=0$ and large enough transverse wavevectors in order to be well within the regime $(\sigma {\bf K}_\perp)^2 \gg 1$, as required for the correspondence between the NSE and the paraxial wave/heat diffusion equation. We isolate the first $\ell = 1$ diffracted order with an iris placed in the Fourier-plane of a 4:1 telescope, which is also used to reduce the beam input ring diameter to $\sim$360 $\mu$m. The OAM beam is input to a lead-doped glass (Schott SF6) plate with dimensions $5\times40\times400$ mm (height, width, length), see Fig.~\ref{fig:expt}. The output facet of the glass medium is then imaged onto an sCMOS camera (Andor Zyla 4.2+). We simultaneously image both the near-field and far-field (spatial-frequency spectrum) intensity profiles of the sample output. The near-field is imaged using an $f_1 = 250$ mm lens providing a magnification of $M = 2.7$. The far-field is imaged by placing an $f_2 = 200$ mm lens at distance $f_2$ from both the output surface and from the camera sensor. The transverse wavevector calibration is given by ${\bf K}_\perp = (x/f_2) k_0$, where $x$ is the transverse position and $k_0 = 2\pi/\lambda$ is the fundamental wavevector. The near and far fields are then monitored as the power of the laser is increased using an attenuator ($\lambda/2$-waveplate followed by a polariser) placed before the phase mask.\\

\noindent {\bf{Data availability.}}  All experimental and numerical data is available at http://dx.doi.org/10.17861/1b1afd55-46dd-49a1-9ed8-3034bd9a63b1.

\section*{Acknowledgements}
D.F. acknowledges financial support from the European Research Council under the European Unions Seventh Framework Programme (FP/2007-"1¤7"1¤72013)/ERC GA 306559 and EPSRC (UK, Grant EP/J00443X/1). N.W acknowledges support from the EPSRC CM-CDT Grant No. EP/L015110/1.

\section*{Author contributions}
All authors contributed to the preparation of the manuscript. The experiments were performed by TR, KW and DV. The theory was developed by EMW, NW and DF. Numerical simulations were performed by CM and EMW. The project was led by DF.\\

\noindent {\bf Competing financial interests:} The authors declare no competing financial interests.

\clearpage

\vspace{30cm}

\noindent {\bf \Large Supplementary Discussion}\\

\noindent We first expand the concept of Boson star rotation in terms of a geometric Berry phase. We then explore the consequence of including the next-to-leading order general relativistic corrections to the Schr\"odinger-Newton equation. Finally, we interpret the regular Schr\"odinger-Newton system as a non-interacting gas coupled to gravity and discuss the behaviour reported in the main text, using this language. \\

\noindent {\bf Analysis of the rotating boson stars in terms of a geometric Berry phase: linear case.}
The rotation of the peaks also occurs to a lesser extent for linear propagation so we look at this first.  For a unit OAM strong beam we write the initial condition as  $(\epsilon <<1)$
\begin{eqnarray}
E(\rho,\theta,z=-L) &=& E_0(\rho)e^{i\theta} \left ( 1+ \epsilon \cos(2\theta) \right )  \\
&=& \underbrace{E_0(\rho)e^{i\theta}}_{\Phi_g(z)} +{\epsilon\over 2}  \underbrace{E_0(\rho)e^{3i\theta} }_{3\Phi_g(z)}+ {\epsilon\over 2}  \underbrace{E_0(\rho)e^{-i\theta} }_{\Phi_g(z)}  \nonumber .
\end{eqnarray}
The three different terms on the lower line have differing OAM $\ell=+1,3,-1$, and therefore accumulate differing Gouy phase-shifts $|\ell|\Phi_g(z)$ through the focus, with $\Phi_g(z)$ the Gaussian beam Gouy phase.  It is the z-dependent phase differential between the $\ell=3,-1$ terms that allows the peaks to rotate.  To see this qualitatively, if one ignores the radial field dependence for simplicity $(\epsilon <<1)$
\begin{equation}
|E(\theta,z)|^2 \simeq 1+2\epsilon\cos(\Phi_g(z))\cos(2\theta + \Phi_g(z))  .
\end{equation}
Now $-{\pi\over 2} < \Phi_g(z) < {\pi\over 2}$, so $\cos(\Phi_g(z))>0$, and the peaks follow $\theta_p(z)=-{\Phi_g(z)\over 2}$ which changes sign through the origin.  (Note that due to the neglect of radial modes this is approximate at best.)

So in the linear case the Gouy phase-shift is at the heart of the rotation of the peaks.  On the other hand it is known that the Gouy phase is a manifestation of a general Berry phase that occurs through a beam focus when the beam parameters undergoe cyclic motion \cite{Sub95,FenWin01}.  The Berry phase is a geometric or topological phase acquired by a system after if is cycled though a closed loop in parameter space.  So the rotation during focusing may be interpreted as a geometric phase effect in the linear case.\\

\noindent {\bf Analysis of the rotating boson stars in terms of a geometric Berry phase: nonlinear case.}
The rotation of the peaks survives in the nonlinear case and appears modified but enhanced due to the nonlinear self-focusing effect.  We cannot offer a detailed analysis of this but we can point to some general expectations that support that the rotation remains a Berry phase effect.  In his paper Subbarao \cite{Sub95} already mentioned that the Berry phase would persist in the Gaussian beam analysis of self-focusing.  Furthermore, Garrison and Chiao \cite{GarChi88} argued that geometrical phases will arise in a general class of nonlinear field theories displaying global gauge invariance, and this applies in our case.  Furthermore their work is not restricted to the adiabatic approximation, so the notion of Berry's phases resulting from nonlinear self-focusing is expected to persist.  Thus the peak rotation in the nonlinear case may be viewed as a nonlinear extension of the linear case whose physical origin lies in the nonlinear Berry's phase during the self-focusing of the fields.\\

\noindent {\bf Effects of General Relativity to the Schr\"odinger-Newton system.} We wish to show that the next-to-lowest order corrections to the NSE, indicate that the full general relativistic description will lead to the similar results as those shown in this work, albeit with a stronger attractive forces. Let us start with deriving the Schr\"odinger-Newton equation from the Klein-Gordon coupled to the weak-field limit of general relativity. In the latter mentioned limit, one can show that Einstein's field equations reduce to
\begin{equation}\label{eq:poisson}
\nabla^2 \Phi = 4\pi Gm\Psi\Psi^{*}
\end{equation}
where the metric tensor components are given by 
\begin{equation}\label{eq:metric}
g_{00} = -(1+2\Phi), \;\;\; g_{ij} = (1-2\Phi)\delta_{ij}, \;\;\; {i,j} = 1,2,3.
\end{equation}
The coupling to the Klein-Gordon equation then follows:
\begin{equation}\label{eq:KleinGordon}
\hbar^2 \left[-(1-2\Phi)\partial^2_{t}+c^2(1+2\Phi)\nabla^2\right]\Psi - m^2c^4\Phi = 0.
\end{equation}
Here we assume that $\Phi \ll 1$ and that $\Phi^2$ is negligible. We can now proceed with the ansatz $\Psi = \psi e^{-i\frac{mc^2}{\hbar}t}$ in eq.~\ref{eq:KleinGordon}. By neglecting the $\mathcal{O}\left(c^{-2}\right)$ term (i.e. slow-moving limit) we arrive at
\begin{equation}
i\hbar\left[1-2\Phi\right]\partial_t \psi = -\left[1+2\Phi\right]\frac{\hbar^2}{2m}\nabla^2\psi+mc^2\Phi\psi
\end{equation}
This is the Schr\"odinger-Newton equation but where the next-to-leading order effects of gravity are taken into account. We can see that by neglecting $\Phi\psi$ products, we get the regular Schr\"odinger-Newton equation. The main difference wth respect to the NSE is that general relativity acts to `amplify' the momentum, that is, $k^2 \rightarrow k^2(1+\Phi)$ when next-to-leading order effects are taken into account. Unstable solutions of the Schr\"odinger-Newton equation are thus unstable also when general relativistic effects are taken into account.\\

\noindent {\bf Schr\"odinger-Newton as a non-interacting gas.} We start from the regular Schr\"odinger equation with a potential $U$. 
\begin{equation}
i\hbar\partial_t \psi = -\frac{\hbar^2}{2m}\nabla^2\psi+U\psi
\end{equation}
It is well know that by a Madelung transform, i.e. $\psi = \sqrt{\rho}e^{iS}$, we can recast this into hydrodynamical equations describing a non-interacting gas:
\begin{equation}
\partial_{t}\rho+\boldsymbol{\nabla}\cdot\left(\rho\vec{v}\right)=0
\end{equation}
describing probability conservation, and
\begin{equation}\label{eq:velocityGeneral}
\partial_t v +\boldsymbol{\nabla}\left[\frac{1}{2}v^2+\frac{U}{m}-\frac{\hbar^2}{2m^2}\frac{\nabla^2\sqrt{\rho}}{\sqrt{\rho}}\right]=0
\end{equation}
describes the `velocity' profile. Note that the `velocity' of the gas is here given by ${\bf v}=\frac{\hbar}{m}\boldsymbol{\nabla} S$. First we can see that there is an outward pressure, usually denoted quantum pressure, given by $P_Q = \frac{\hbar^2}{2m^2}\frac{\nabla^2\sqrt{\rho}}{\sqrt{\rho}}$, which is relevant when the density grows rapidly. \\
\indent Now we let this `gas' gravitate. Or in other words, we let $U = mc^2\Phi$, where $\Phi$ is determined by eq.~\ref{eq:poisson}. Substituting this into eq.~\eqref{eq:velocityGeneral}, we find
\begin{equation}\label{eq:velocityGravitating}
\partial_t v +\boldsymbol{\nabla}\left[\frac{1}{2}v^2+c^2\Phi-\frac{\hbar^2}{2m^2}\frac{\nabla^2\sqrt{\rho}}{\sqrt{\rho}}\right]=0
\end{equation}
which is coupled to 
\begin{equation}\label{eq:gravity}
\nabla^2 \Phi = 4\pi Gm\rho.
\end{equation}
We note that in the present context $\rho$ is \textit{not} a mass density, but a probability density. \\ \newline
{\bf Conclusions drawn from the gas picture.} The above equations are very similar to the hydrodynamical equations commonly used when discussing star formation. However, there are some important differences. The gas we see above is non-interacting. This implies that there is no direct pressure arising from `particle' bunching, which is required for normal hydrostatic equilibrium. Hydrostatic equilibrium is however still possible, but gravity now needs to be balanced by the quantum pressure $P_{Q}$. Boson stars are indeed stable due to the Heisenberg uncertainly principle \cite{liebling2012dynamical}, and here we see that this manifests itself naturally through the quantum pressure term in the hydrodynamical picture.\\ 
\indent We now focus attention on the implications of the fact that for a wavefunction with angular momentum, there is a phase singularity at the centre. At this central point, the hydrodynamical picture is not well defined. However, we can see the effective probability `gas' will naturally avoid this point, as there the term $\frac{1}{2}v^2 \rightarrow \infty$ in eq.~\eqref{eq:velocityGravitating}. The centre is thus a point of infinite repulsive pressure, which is just another description of a perfect mirror. Another viewpoint is that this point has infinite potential energy, as we see that it enters in the same way as the gravitational potential $\Phi$ in eq.~\eqref{eq:velocityGravitating}. Either way, our conclusions remain the same: any `particles' that encounter the centre will bounce away from it.\\

\end{document}